\documentclass[prl,superscriptaddress,showpacs,twocolumn]{revtex4}

\usepackage{amsmath}
\usepackage{amsfonts}
\usepackage{amssymb}
\usepackage{graphicx}

\newcommand{\beq}{\begin{equation}}
\newcommand{\eeq}{\end{equation}}
\newcommand{\smallfrac}[2]{\mbox{$\frac{#1}{#2}$}}
\newcommand{\half}{\smallfrac{1}{2}}
\newcommand{\ip}[2]{\left\langle{#1}|{#2}\right\rangle}
\newcommand{\sq}[1]{\left[ {#1} \right]}
\newcommand{\cu}[1]{\left\{ {#1} \right\}}
\newcommand{\ro}[1]{\left( {#1} \right)}
\newcommand{\pr}[1]{\operatorname{Pr}\sq{#1}}

\newcommand{\ket}[1]{\ensuremath{\left| #1 \right\rangle}}
\newcommand{\bra}[1]{\ensuremath{\left\langle #1 \right|}}

\newcommand{\C}[1][{}]{\ensuremath{C^\text{#1}}}

\newcommand{\sysa}[1][\pm]{\ensuremath{\ket{\psi_{#1}}}}
\newcommand{\sysp}{\sysa[+]}
\newcommand{\sysm}{\sysa[-]}
\newcommand{\sysap}[1][a]{\ensuremath{\ket{\psi_{#1}^\perp}}}
\newcommand{\syspp}{\sysap[+]}

\begin{document}
\title{Mixed state discrimination using optimal control}

\author{B.~L. Higgins}
\affiliation{Centre for Quantum Dynamics, Griffith University, Brisbane, 4111, Australia}
\author{B.~M. Booth}
\affiliation{Centre for Quantum Dynamics, Griffith University, Brisbane, 4111, Australia}
\author{A.~C. Doherty}
\affiliation{Physics Department, The University of Queensland, Brisbane, 4072, Australia}
\author{S.~D. Bartlett}
\affiliation{School of Physics, The University of Sydney, Sydney, 2006, Australia}
\author{H.~M. Wiseman}
\email{H.Wiseman@griffith.edu.au}
\affiliation{Centre for Quantum Dynamics, Griffith University, Brisbane, 4111, Australia}
\author{G.~J. Pryde}
\email{G.Pryde@griffith.edu.au}
\affiliation{Centre for Quantum Dynamics, Griffith University, Brisbane, 4111, Australia}

\begin{abstract}
We present theory and experiment for the task of discriminating two nonorthogonal states, given multiple copies. We implement several local measurement schemes, on both pure states and states mixed by depolarizing noise. We find that schemes which are optimal (or have optimal scaling) without noise perform worse with noise than simply repeating the optimal single-copy measurement. Applying optimal control theory, we derive the globally-optimal local measurement strategy, which outperforms all other local schemes, and experimentally implement it for various levels of noise.
\end{abstract}


\maketitle

Quantum control---the application of control theory to quantum systems---offers powerful tools to enable quantum technologies to function robustly in the presence of noise and device imperfections~\cite{WisMil09,Armen_PRL_89_2002,Geremia03,Branczyk2007,JM07}, and to simplify protocols by reducing the need for entangling operations or collective measurements~\cite{Griffiths1996,Walgate2000}. One such tool is adaptive measurement, wherein one adapts future measurements based on the outcomes of previous ones~\cite{WisMil09}. Quantum control based on adaptive measurements has been used to improve the measurement of an optical phase~\cite{Armen_PRL_89_2002,Higgins_Nature_450_2007,Higgins2009}. Here, we consider the problem of quantum state discrimination, and demonstrate experimentally that adaptive local measurements can discriminate pure states better than nonadaptive ones. Moreover, we show that in the presence of noise, which is unavoidable in practice, the full power of optimal control theory is required to derive the globally-optimal adaptive (local) measurement scheme, which we then experimentally implement.

The task of state discrimination is a fundamental primitive in many fields of quantum information science, including quantum communications, cryptography, and computing. If a quantum system is prepared in one of several possible states, this preparation can only be determined with certainty if the possible states are all mutually orthogonal. For nonorthogonal states, two complementary tasks are often considered~\cite{WisMil09}: minimizing the likelihood of either an incorrect result (an error)~\cite{Helstrom1976}, or of an inconclusive result with no errors~\cite{Ivanovic1987,Dieks1988,Peres1988}.

In this Letter, we consider the minimum-error discrimination of two nonorthogonal qubit states, given $N$ identical copies of the state, using only local measurements, where the cost function $C_N$ (which is to be minimized) is the probability of error. While continuous measurement schemes for distinguishing two infinite-dimensional pure states from a single copy have been studied elsewhere~\cite{JM07,Wittmann2008}, here we consider discrete measurements of each of $N$ discrete copies of the state. An optimal solution for multiple-copy discrimination of pure states is given by Helstrom~\cite{Helstrom1976} (see also~\cite{WisMil09}), and takes the form of a two-outcome projective measurement on the joint space of all copies. For $N>1$, this measurement is a nonlocal (collective) measurement on all copies, and schemes in which the same local measurement is performed on each system do not achieve this optimal performance~\cite{Acin2005}. Remarkably, it has been predicted theoretically that the optimum can be reached using \emph{adaptive} local measurements~\cite{Acin2005}. In this adaptive scheme each system is measured locally in the basis that minimizes the probability of error immediately after that measurement. We refer to this procedure of $N$ adaptive measurements as the ``locally-optimal local measurement'' scheme. As shown in~\cite{Acin2005}, for pure states this adaptive measurement performs just as well as the optimal collective measurement on all $N$ copies of the state. In the asymptotic limit $N \rightarrow \infty$, the scaling of $C_N$ for various state discrimination schemes has been well studied~\cite{Acin2005,Calsamiglia2008,Hayashi2008}, with the notable finding that adaptive local measurements do not provide an advantage (in terms of scaling) over fixed strategies, even for mixed states~\cite{Hayashi2008}.

Although the asymptotic performance of state discrimination schemes is of considerable academic interest, practical applications will require results for finite $N$, and moreover must consider the effect of noise (i.e.\ mixed states). Here, we adjust the local measurement strategies presented in Ref.~\cite{Acin2005} to function in the presence of noise, and analyze their performance theoretically and experimentally. Importantly, we discover that, with the exception of states that are almost pure, simple nonadaptive ``unbiased measurements'' (see below) outperform the locally optimal strategy defined above, for a sufficiently large number of copies. However, the globally-optimal local measurement strategy, determined using optimal control theory, does outperform unbiased measurements, even though it does not achieve the optimum achievable using nonlocal measurements. For $N$ up to 10, we theoretically predict and experimentally demonstrate the performance of each scheme with various levels of noise.

All measurements we consider are projective, in a basis $\cu{\ket{\phi},\ket{\phi-\pi/2}}$, where $\phi \in \left[ 0,\pi/2 \right)$ and $\ket{\phi} \equiv  \cos \phi  \ket{x}+\sin \phi  \ket{y}$, for some orthonormal basis $\{\ket x$, $\ket y\}$. Initially, we restrict our study to the problem of distinguishing between two nonorthogonal pure states, defined without loss of generality by $\sysa = \cos\theta\ket x \pm \sin\theta\ket y$. Their overlap is $c = \ip{\psi_+}{\psi_-} = \cos 2\theta$, and they are prepared with probability $q_\pm$ ($q_+ \geq q_-$). The single-copy Helstrom measurement is the projective measurement with $\phi^\text{Hel}(q_+) = \frac{1}{2}\operatorname{arccot}\left(\left(q_+ - q_-\right)\cot2\theta\right)$. From a measurement on a single copy, the most likely state given outcome $+$ ($-$) is $\sysp$ ($\sysm$), and the probability of error resulting from this best guess is $\C[Hel]_1 = (1 - \sqrt{1 - 4 q_+ q_- c^2})/2$. 

For multiple copies, we first build upon the three local measurement schemes presented in Ref.~\cite{Acin2005}. We treat these schemes as a prescription for what measurements to make, but unlike~\cite{Acin2005} we employ Bayesian processing of all results. This analysis allows us to determine the performance of these schemes for distinguishing mixed states. For pure states, however, such analysis is equivalent to the protocols as presented in Ref.~\cite{Acin2005}. 

{\em 1.\ Unbiased measurements}:
Independently perform the single-copy Helstrom measurement on each copy, and decide in favor of the state with the highest posterior probability. For pure states with $q_+ = q_-$, this decision reduces to choosing the state with the most favorable outcomes---a ``majority vote'' as in Ref.~\cite{Acin2005}. When $N$ is even there is the potential of a ``split vote'', in which case a random guess is made. This scheme performs for general states and odd $N$ as $\C[un]_N = \sum_{m>N/2}^{N} \binom{N}{m} (\C[Hel]_1)^m (1 - \C[Hel]_1)^{N-m}$, and $\C[un]_N = \C[un]_{N-1}$ for even $N$. For pure states, the large $N$ scaling is $\C[un]_N \sim \eta c^N$, where $\eta$ is a constant.

{\em 2.\ Fully biased measurements}:
Independently perform a projective measurement on each copy with $\phi=\theta$; that is, in the basis $\{\sysp, \syspp\}$. For pure states, the scheme can only guess the $\sysp$ state if all measurement results are $\sysp$, otherwise it must guess $\sysm$; this is the ``unanimity vote'' scheme of Ref.~\cite{Acin2005}. Mixed states, however, cannot reliably fulfill unanimity; in general, the best guess must be made via Bayesian analysis. For pure states, the error probability is $\C[fb]_N = q_+ c^{2N}$. Asymptotically, this scheme thus scales quadratically better than unbiased measurements; however, when $N$ is sufficiently small, unbiased measurements have better performance.

{\em 3.\ Locally-optimal local measurements}:
Perform an optimal single-copy Helstrom measurement $\phi^\text{Hel}(q_+)$ on the first copy. Via Bayes' theorem, use the result to update the prior probability $P_1 = q_+$ to posterior probability $P_2$. Using this, apply a new single-copy Helstrom measurement $\phi^\text{Hel}(P_2)$ on the next copy. Repeat this adaptive process with updated probabilities $P_n$ for all remaining copies. The best guess is the state with the higher final posterior probability. For pure states, this scheme is globally optimal for all $N$~\cite{Acin2005}, yielding the same probability of error as the collective $N$-copy Helstrom measurement: $\C[loc]_N = (1 - \sqrt{1 - 4 q_+ q_- c^{2N}})/2 \sim q_+ q_- c^{2N}$.

\begin{figure}[tb]
\center{\includegraphics[width=6.88cm]{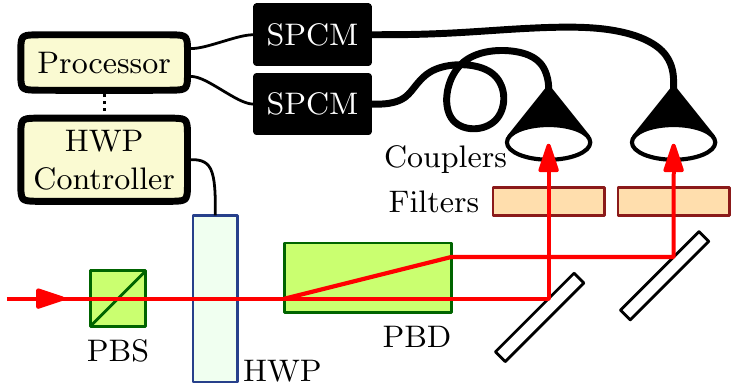}}
\caption{Layout of the experiment. A polarizing beam splitter (PBS) acts as a filter to ensure high fidelity horizontally-polarized photons. A half-wave plate (HWP) in a motorized rotation stage determines the measurement basis. A polarizing beam displacer (PBD) and single-photon counting modules (SPCMs) discriminate between horizontally and vertically polarized photons with high contrast. The result of the measurement is fed to a processor which, depending on the protocol being tested, adjusts the operation of the HWP controller. Single photon inputs are obtained through type-I spontaneous parametric downconversion---a $410 \text{ nm}$ diode laser pumps a BiBO (bismuth borate) crystal, producing pairs of $820 \text{ nm}$ single photons in the state $\ket{HH}$, with photons in separate spatial modes. One of the photon pair is guided to the input of the experiment through a single-mode optical fibre. The other photon is guided directly to a single-photon counting module. Detection in coincidence ensures high fidelity single-photons are measured in the experiment.}
\label{fig:expt}
\end{figure}
 
We experimentally demonstrate these schemes with $\theta = 15^\circ$ and $q_+ = q_- = 1/2$, using single photon polarization to encode the two pure states we wish to discriminate; see Fig.~\ref{fig:expt}. Within the experiment, horizontal photon polarization implements the $\ket x$ and vertical polarization implements the $\ket y$ basis states. A half-wave plate (HWP) determines the measurement basis. The measurement outcomes are entirely dependent on the relative angle between the state and the measurement axes, and not on any global orientation of the state or measurement axes. Therefore, we do not separately prepare the two states $\sysp$ and $\sysm$, but rather always prepare $\sysp$ and offset the measurement axes by an angle $2\theta$ for experiments on $\sysm$. A high-contrast-ratio polarizing beam displacer and single photon counting modules implement the orthogonal measurement outcomes. The polarization contrast ratio achievable with the apparatus was measured to be better than $0.9999$ (the Bayesian processing assumes perfect visibility). The results of running each of the three algorithms in the experiment, and their theoretical predictions, are shown in Fig.~\ref{fig:nonoise}---the experimental results correspond well with the theoretical predictions.

\begin{figure}[tb]
\center{\includegraphics[width=8.5cm]{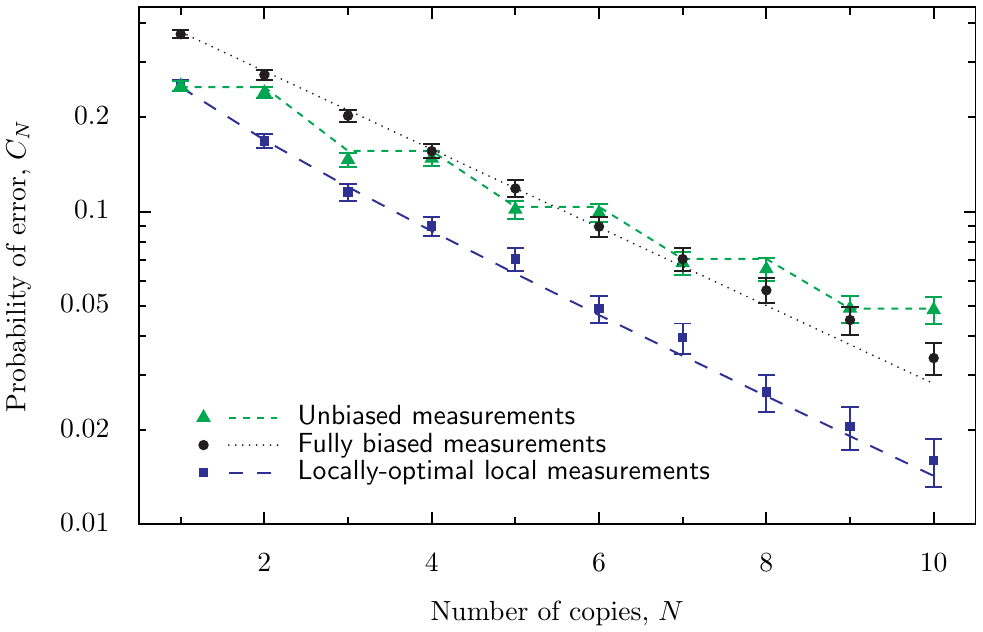}}
\caption{Probability of error $C_N$ in $N$-copy state discrimination using various schemes in the absence of noise. Lines represent theoretical predictions; points represent experimental data, each $2000$ measurements. Error bars represent one standard deviation of the mean of a binomial distribution. The locally-optimal local measurement scheme performs best for all $N$; where $N=1$ it is equivalent to unbiased measurements, both schemes using one single-copy optimal measurement.}
\label{fig:nonoise}
\end{figure}

We now turn to the performance of these schemes in the presence of noise, i.e., for mixed states. This situation describes the addition of noise due to, for example, transmission over a noisy channel, as well as imperfect measurements. In particular, we consider uniform depolarizing noise on qubits~\cite{Nielsen2000} of strength $0\leq \nu \leq 1$, so that the two states are now
\beq
\rho_\pm = \half [1 + (1-\nu)(Z \cos 2\theta \pm X \sin 2\theta)].
\eeq
Here $X = \ket{x}\bra{y}+\ket{y}\bra{x}$, and $Z = \ket{x}\bra{x} - \ket{y}\bra{y}$ are Pauli operators. This is simulated experimentally by performing bit, phase, and bit-phase flips in the measurement basis, each with probability $\nu/4$. Because the noise is depolarizing, the angles for the fixed measurement schemes are the same in the mixed state case as in the pure case. Even so, noise will clearly have a detrimental effect on the performance of the schemes described above. Indeed, it is now the case that no local scheme can achieve the globally optimal performance achievable with a collective measurement.

We have calculated the respective error probabilities $C_N$ exactly as a function of noise; see Figs.~\ref{fig:varn} and \ref{fig:varnoise}. Both the fully biased measurement scheme and the locally-optimal adaptive scheme lose their superiority over  unbiased measurements as $\nu$ is increased. Our theoretical analysis confirms this behavior for general $\theta$ and $q_+$, with the value of $\nu$ at which the error probability curves cross depending on $\theta$, $q_+$, and $N$.

The locally optimal scheme maximizes the discriminating power of each measurement individually, but that is not the same as maximizing the discriminating power of all $N$ measurements together (even when restricting to local, not collective, measurements). Because the locally-optimal local measurement scheme is evidently not the globally-optimal local measurement scheme \emph{in general}, we now turn to finding such a scheme.

{\em 4.\ Globally-optimal local measurements}:
To determine the optimal discrimination scheme using local adaptive measurements, we use dynamic programming~\cite{Nemhauser1966}. This will in general yield an adaptive scheme that depends explicitly on the total number of measurements $N$ that will be performed, unlike the locally optimal scheme. The scheme is defined by a table of measurement angles, with rows corresponding to $n$, the copy to be measured ($1 \leq n \leq N$), and columns corresponding to $P_{n}$, the probability prior to the $n$th measurement that the prepared state is $\sysp$, conditioned on the measurement results of the first $n-1$ copies ($P_1 = q_+$). Thus, at the $n$th step, we consult the table to obtain the measurement angle $\phi_{n}(P_{n})$ to be used. The result of this measurement is then used to calculate a posterior $P_{n+1}$ via Bayes' theorem, and we proceed to the next step. Linear interpolation resolves the discreteness in the table's representation of $P_{n}$ (here we use $2501$ samples).

We construct this table as follows. In all cases, the optimal measurement on the final copy $n=N$ must be the single-copy Helstrom measurement, $\phi_{N}(P_{N}) = \phi^\text{Hel}(P_{N})$, as this measurement will minimize the error probability $C_N$ regardless of the previous measurement choices.  Starting from this fact, the globally-optimal local measurement scheme for $N$ copies is constructed in reverse. Using the recursive relationship between the expected error probabilities after $n$ and $n+1$ measurements, the penultimate measurement $\phi_{N-1}(P_{N-1})$ that minimizes $C_N$ can be found by a numerical search, given $P_{N-1}$. When calculated for samples of the range $0 \leq P_{N-1} \leq 1$, this defines row $N-1$ of the measurement table.

The optimal measurement that precedes the final two measurements can similarly be obtained by minimizing the expected error probability over the measurement $\phi_{N-2}(P_{N-2})$ for some $P_{N-2}$. This constructs row $N-2$ of the measurement table. Continuing this analysis, we construct a table of $N$ measurement settings, defining $\phi^\text{glo}_{n}(P_{n})$, which results in the lowest final error probability, $\C[glo]_N$.

To determine the performance of non-globally-optimal measurements when noise is present one can use the same procedure, but with a nonoptimal measurement choice. For example, the locally-optimal local measurement $\phi_{n}^\text{loc}(P_{n})$ defines  $\C[loc]_N$. As the measurements are Markovian, sampling is unnecessary, and for moderate $N$ as used in this paper, the probability of error can be calculated exactly.

The globally-optimal local measurement scheme constructed according to the above procedure reduces to the locally optimal scheme in the noiseless case. For high noise, we have found numerically that the measurement setting $\phi^\text{glo}$ for all but the final few copies approaches $\pi/4$, as for unbiased measurements when $q_+ = q_-$. Its performance also approaches that of unbiased measurements, in this regime where $\nu$ is not small. Importantly, for all $\nu>0$, we have $\C[glo]_N < \min(\C[loc]_N, \C[un]_N)$ for $N\geq 3$, as expected. But we also have $\C[glo]_N > \C[col]_N$, the probability of error from a collective measurement over all copies, achieved by the $N$-copy Helstrom measurement~\cite{Helstrom1976,WisMil09}.  This is illustrated in Figs.~\ref{fig:varn} and \ref{fig:varnoise}.

\begin{figure}[tb]
\center{\includegraphics[width=8.5cm]{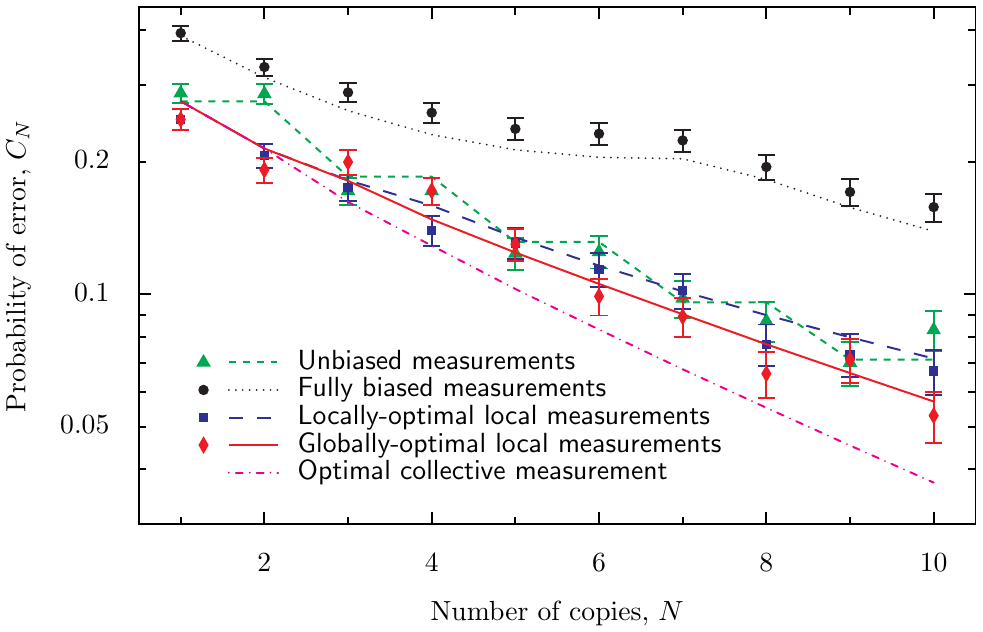}}
\caption{Error probability $C_N$ of discrimination schemes under $\nu = 10\%$ depolarizing noise. Points represent $1000$ experimental discriminations. The addition of noise detrimentally impacts the locally-optimal local measurement scheme more than the unbiased scheme. Indeed, theory predicts that the latter outperforms the former for $N = 5$, $N = 7$, and $N \geq 9$. The globally-optimal local measurement scheme performs better than all other local measurement schemes in the presence of noise. The theoretical optimal collective measurement cost is plotted for comparison.}
\label{fig:varn}
\end{figure}

\begin{figure}[bt]
\noindent\center{\includegraphics[width=8.5cm]{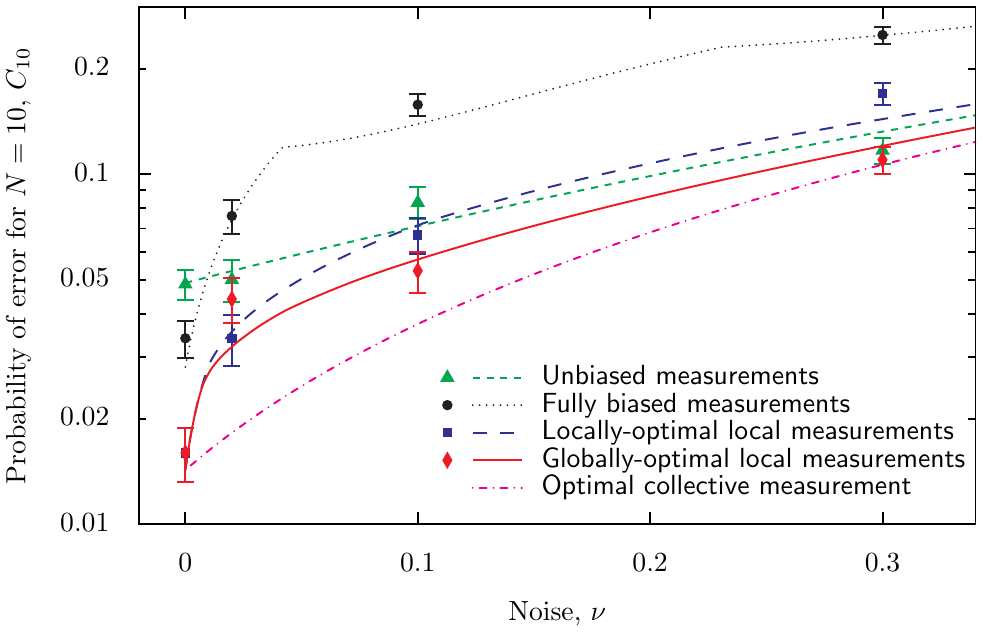}}
\caption{Error probability $C_N$ of discrimination schemes under various levels of noise $\nu$ for $N=10$ measured copies. Points each represent $2000$ ($\nu=0$) or $1000$ ($\nu > 0$) experimental discriminations. Here, unbiased measurements outperform the locally-optimal local measurement scheme for noise $\nu \gtrsim 10\%$.}
\label{fig:varnoise}
\end{figure}

We experimentally investigate all four local measurement schemes in the presence of 2\%, 10\%, 30\%, and 60\% noise. The results for $\nu=10\%$ noise for $N$ up to 10 are shown in Fig.~\ref{fig:varn}, and for fixed $N=10$ under various noise in Fig.~\ref{fig:varnoise}. Further results may be found in supplementary material, below. Theoretical curves are determined numerically using the dynamic programming method described above. The discontinuities in the gradient of $\C[fb]_N$ arise due to the discreteness of the number of outcomes required to guess $\sysm$. In all cases, experimental data agree with theoretical predictions, within expected statistical variation. The globally optimal scheme has the best performance for all levels of noise and for all $N$.

We have shown that local adaptive $N$-copy discrimination schemes which are optimal in the noiseless regime are significantly impacted by the addition of noise. The locally-optimal local measurement scheme, in particular, performs more poorly than nonadaptive unbiased measurements. Subsequently, by a dynamic programming analysis, we have demonstrated the adaptive local measurement scheme that is globally optimal, having, in all cases, the lowest probability of an incorrect discrimination of any local measurement scheme for any $N$. In addition to illuminating part of the fundamentally interesting problem of quantum state discrimination, our work provides an insight into the fragility of idealized models practically applied, and demonstrates the usefulness of optimal quantum control techniques in mitigating the real-world issues that face the application of quantum technologies.

We thank Ramon Mu\~noz-Tapia for helpful discussions. This work was supported by the Australian Research Council.

\begin{titlepage}

\center{\textbf{\begin{large}Supplementary material to\\
Mixed state discrimination using optimal control\end{large}}}\\

\center{B.~L. Higgins,$^1$ B.~M. Booth,$^1$ A.~C. Doherty,$^2$ S.~D. Bartlett,$^3$ H.~M. Wiseman,$^1$ and G.~J. Pryde$^1$}

\vspace{-0.5em}
\center{\begin{small}\textit{$^1$Centre for Quantum Dynamics, Griffith University, Brisbane, 4111, Australia\\
$^2$Physics Department, The University of Queensland, Brisbane, 4072, Australia\\
$^3$School of Physics, The University of Sydney, Sydney, 2006, Australia}\end{small}}\\

\vspace{38pt}

\end{titlepage}

Here we present additional details and results of $N$-copy discrimination schemes for two non-orthogonal quantum states under depolarizing noise.

\section{Noise Simulation}

Single-qubit depolarising noise is simulated in the experiment by the random application of bit, phase, and bit-phase flips in the measurement basis, each with probability $\nu/4$, where $0 \leq \nu \leq 1$ quantifies the amount of noise. For each copy, the appropriate measurement, described by the angle $\phi$, is first calculated according to the scheme being tested. Before passing to the half-wave plate controller, this angle is passed through a depolarising filter subroutine, isolated from the main discrimination routines. The subroutine will perform the operation $\phi \rightarrow \phi$ with probability $1 - \frac{3}{4}\nu$, or $\phi \rightarrow \pi/2 - \phi$, $\phi \rightarrow - \phi$, or $\phi \rightarrow \pi/2 + \phi$, each with probability $\nu/4$, implementing identity, bit, phase, and bit-phase flip operations respectively. This realizes a noisy measurement equivalent to a depolarising channel of strength $\nu$.

\section{Derivation of Globally-Optimal Local Measurements}

The table of measurements that defines the globally-optimal local measurement scheme for $N$ copies is constructed as follows. Let $P_{n+1}$ be the probability (i.e.\ the observer's credence) that the prepared state is $\sysp$, conditioned on the $n$ measurement results from the first $n$ copies. Let $R_n^\text{glo}$ be the expected value (calculated after the $n$th measurement) of the final probability of error after measuring the remaining $N-n$ copies using globally-optimal local measurements. Let $\phi_{n}$ be a parameter defining the measurement basis for the $n$th measurement, and $D_n$ be its outcome. We begin with the condition that, after all copies have been measured, $R_N^\text{glo}(P_{N+1}) = \min \ro{P_{N+1}, 1 - P_{N+1}}$, and proceed iteratively in reverse. Given $R_n^\text{glo}(P_{n+1})$ for some $n>0$, and measurement angle $\phi_{n}$, it is evident that at the previous step the final error probability $R_{n-1}(P_n,\phi_n)$ after measuring the $n$th copy with angle $\phi_{n}$ and the remaining $N-n$ copies using globally-optimal local measurements is
\begin{eqnarray}
&&R_{n-1}(P_{n},\phi_{n}) \\
&&= \sum_{D_n} \pr{D_n | P_{n}, \phi_{n}} R_n^\text{glo} \ro{P_{n+1} \ro{D_n, P_{n},\phi_{n}}}, \nonumber
\end{eqnarray}
where we use Bayes' theorem to evaluate
\begin{equation}
P_{n+1}(D_n,P_{n},\phi_{n}) = \frac{\pr{D_n | +,\phi_{n}} P_{n}}{\pr{D_n | P_{n}, \phi_{n}} }.
\end{equation}
Here $\pr{D_n | P_{n}, \phi_{n}} = \pr{D_n | +,\phi_{n}} P_{n} + \pr{D_n | -,\phi_{n}} (1 - P_{n})$.
The globally optimal measurement at step $n-1$ is defined by finding the angle $\phi_{n}^\text{glo}(P_{n})$ that minimizes $R_{n-1}$, and this defines
\beq
R^\text{glo}_{n-1}(P_{n}) \equiv R_{n-1}(P_{n},\phi_{n}^\text{glo}(P_{n})).
\eeq

This process is then continued down to $n=1$.  The probability of error for this scheme is thus $\C[glo]_N = R^\text{glo}_0(q_+)$ (since $P_1=q_+$). Once this analysis is completed, the values stored in $\phi_n^\text{glo} \ro{P_n}$ define the measurements to be performed within the experiment.

\section{Additional Results}

Following are plots of the error probability of unbiased measurements, fully biased measurements, locally-optimal local measurements, globally-optimal local measurements, and optimal collective measurements, under various levels of depolarizing noise $\nu$, for $N$ up to 10. Shown are theoretical calculations (lines) and experimental data for 1000 discriminations (points, local schemes only), with error bars plus or minus one standard deviation of the mean. In all cases, the globally-optimal local measurement scheme, constructed using optimal control theory and dynamic programming as detailed above, has the best performance of any local measurement scheme for any $N$.

\vspace{0.6em}
\vspace{\stretch{1}}

\begin{minipage}{\linewidth}
{\center\includegraphics[width=8.5cm]{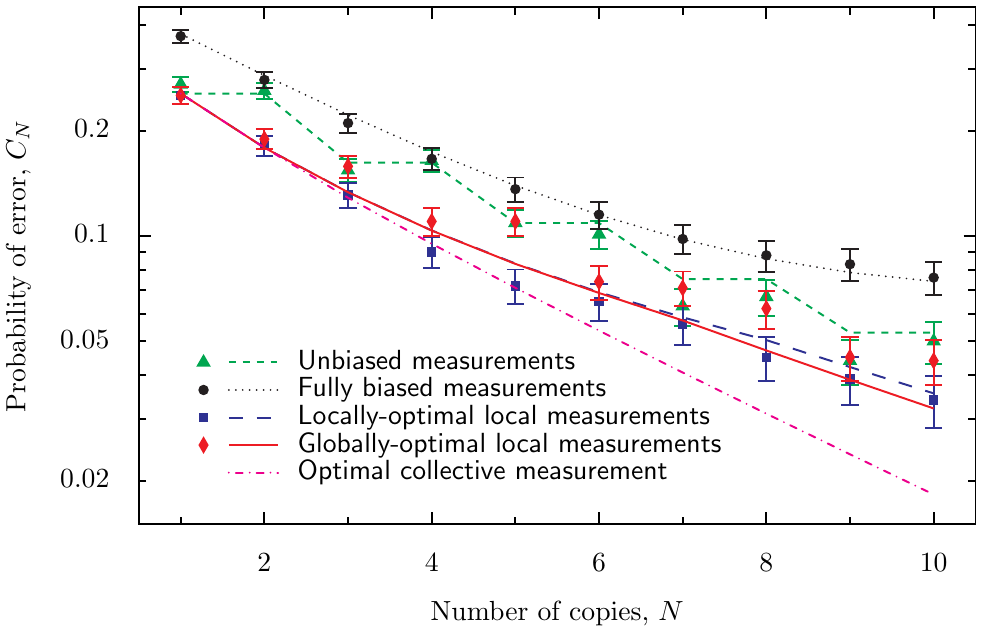}}
{\par\noindent\small\rmfamily FIG.\ 1: Error probability $C_N$ of discrimination schemes under $\nu = 2\%$ depolarizing noise.}
\vspace{2em}
\end{minipage}

\vspace{\stretch{1}}

\begin{minipage}{\linewidth}
{\center\includegraphics[width=8.5cm]{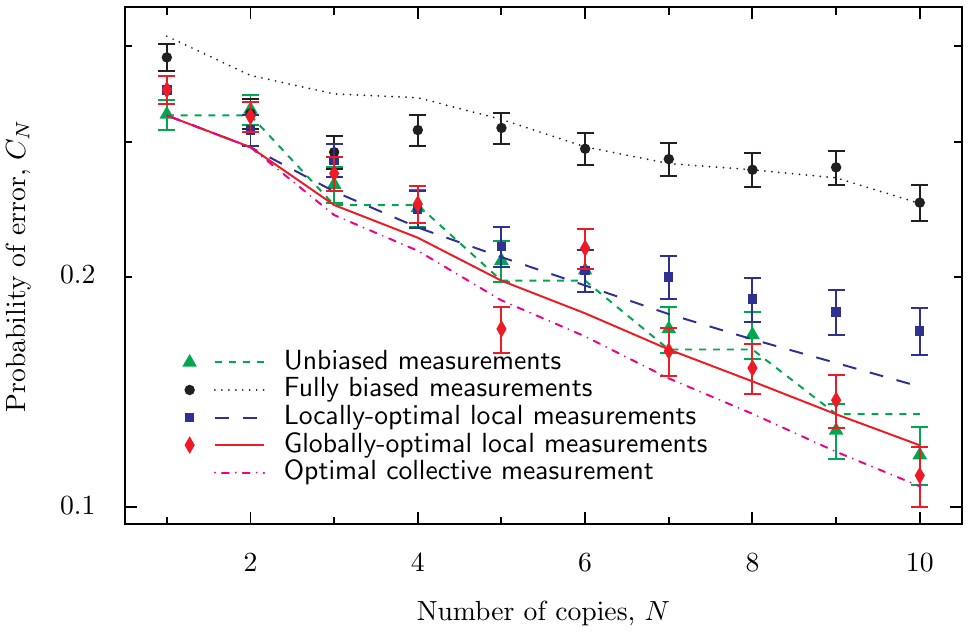}}
{\par\noindent\small\rmfamily FIG.\ 2: Error probability $C_N$ of discrimination schemes under $\nu = 30\%$ depolarizing noise.}
\vspace{2em}
\end{minipage}

\vspace{\stretch{1}}

\begin{minipage}{\linewidth}
{\center\includegraphics[width=8.5cm]{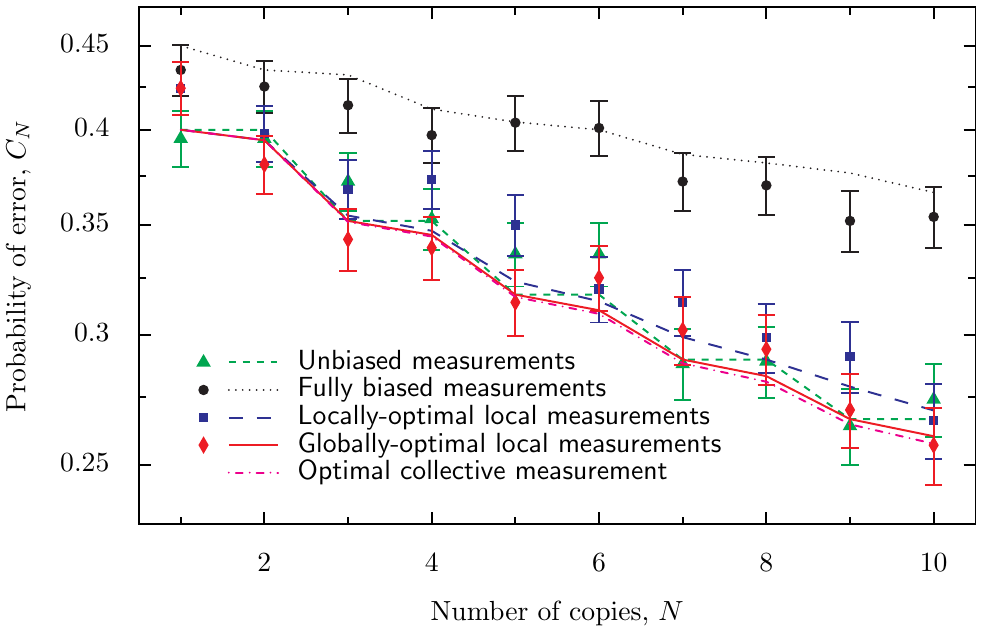}}
{\par\noindent\small\rmfamily FIG.\ 3: Error probability $C_N$ of discrimination schemes under $\nu = 60\%$ depolarizing noise.}
\vspace{5em}
\end{minipage}

\end{document}